\documentclass[prd,nofootinbib,onecolumn,english]{revtex4}
\usepackage{amsmath}
\usepackage{amsfonts,color}
\usepackage{tensor}
\usepackage{amssymb,float}
\usepackage[utf8]{inputenc}
\usepackage{color}
\usepackage[unicode=true,pdfusetitle,
 bookmarks=true,bookmarksnumbered=true,bookmarksopen=true,bookmarksopenlevel=3,
 breaklinks=false,pdfborder={0 0 1},backref=false,colorlinks=true]
 {hyperref}
\usepackage{enumitem}
\usepackage{graphicx}
\usepackage{appendix}
\usepackage{hyperref}
\usepackage{lipsum}
\usepackage{multirow}

\newcommand{\hubble} {Hubble-Lema\^{i}tre\,}

\begin{document}

\title{A dynamical dark energy solution to Hubble tension in the light of the multimessenger era}

\author{Celia Escamilla-Rivera}
\email{celia.escamilla@nucleares.unam.mx}
\affiliation{Instituto de Ciencias Nucleares, Universidad Nacional Aut\'onoma de M\'exico, 
Circuito Exterior C.U., A.P. 70-543, M\'exico D.F. 04510, M\'exico.}


\begin{abstract}
We show that the gravitational waves measurements have raised the opportunity to measure $H_0$ with dark sirens to within 2$\sigma$, the accuracy required to resolve the \hubble tension. There are two principal reasons for our results: (1) upgrades to GW LIGO-Virgo transient catalogues GWTC-1 and GWTC-2 enhance their sensitive with only 10\% of contamination fraction, and (2) new dark sirens should help to constrain our dynamical EoS. In conjunction, sensitivity upgrades and a new dark energy model will facilitate an accurate inference of the \hubble constant $H_0$ to better with an $\pm 0.077$ error in comparison to the LIGO dark siren with $+14.0$/$-7.0$, which would further solidify the role of dark sirens in late dark energy for precision cosmology in the future.
\end{abstract}

\maketitle
 
\section{Introduction}

\begin{figure*}
\centering
\includegraphics[width=0.65\textwidth]{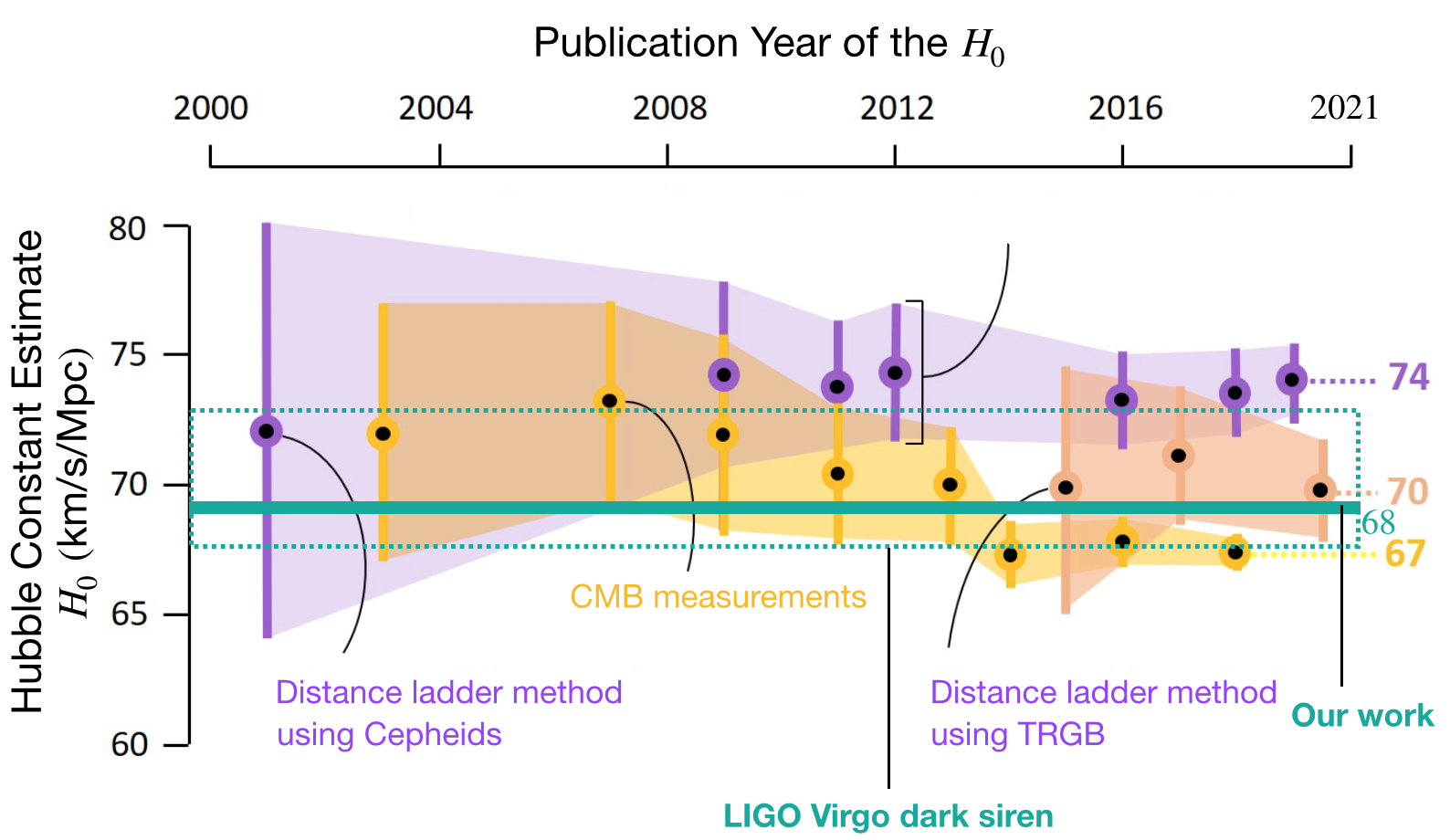}
\caption{Hubble Constant estimations according to the publication year. Planck 2018-based, early universe mean value for $H_0$ is 67 km/s/Mpc.
    The Cepheid-based, late universe mean value is 74 km/s/Mpc. A new alternative to Cepheids, knows as red giant stars that flare with a known intrinsic brightness, only complicated the tension. They indicated a $H_0$ mean value of about 70 km/s/Mpc, a value in the midway between the other two, but in agreement with both the measurements within $2\sigma$. Also, we indicated the independent estimation from the multi-messenger detection of a binary neutron star inspiral $H_0=70.0^{+12.0}_{-8.0}$ km s$^{-1}$ Mpc$^{-1}$. Dark siren detected in the first and second observing runs of LIGO and Virgo estimated  $H_0=68.0^{+14.0}_{-7.0}$ km s$^{-1}$ Mpc$^{-1}$.
    As we shall discuss in Sec.\ref{sec:results}, our proposal allows to determine a value of $H_0=68.0\pm 0.077$ km s$^{-1}$ Mpc$^{-1}$ using GW transient catalog solely.}
\label{fig:H0-all}
\end{figure*}

The \hubble constant $H_0$ is a fundamental cosmological parameter that governs the expansion rate of our Universe.  
However, since the first release of the Cosmic Microwave Background (CMB) observations by the Planck Collaboration in 2013~\cite{Ade:2013zuv}, the determination of $H_0$ based on the standard model of cosmology $\Lambda$CDM, which started to be in tension\footnote{Let us consider two different measurements of the same parameter A. Let $A_1= a_1\pm b_1$, be the first measurement and $A_2 = a_2\pm b_2$, the second measurement. With $a_1$, $a_2$ the mean values and $b_1$, $b_2$ the standard deviation. If we assume that these measurements are independent and that the uncertainties have approximately a normal distribution, then the \textit{tension} between $A_1$ and $A_2$ is given by: $T=|a_1 -a_2|/(b_{1}^{2} +b_{2}^{2})^{1/2} \sigma$. This $T$ quantity indicates the number of standard deviations by which $A_1$ and $A_2$ differ.} with the model-independent determination via calibrated local Supernovae Ia (SNeIa) by the Hubble Space telescope in 2011~\cite{Riess_2011}. An initial tension of around 2.4$\sigma$ has been the worst over the last years. Attached to this, we are dealing with a more troublesome issue: we are able to perform very precise measurements of $H_0$ in several independent ways, but such methods do not agree among each other at a statistically significant level.
This intense debate have led us to two ways to determine
the \hubble constant: one established by the Planck Collaboration in 2018~\cite{Aghanim:2018eyx} assuming the $\Lambda$CDM model, and the other by the SH0ES Collaboration in 2019~\cite{Riess:2019cxk}. Up to now, $H_0$ inferred\footnote{We say \textit{inferred}, because the determination of a $H_0$ value directly relies on the assumption of a cosmological model (in the method employed by \textit{Planck} it is a standard $\Lambda$CDM model), and is strongly connected with early cosmic times physics (prior to recombination).} from the fluctuation spectrum of the CMB ($H_0=67.36\pm0.54$ km s$^{-1}$ Mpc$^{-1}$) disagrees with the value obtained from the measurement of the luminosity distance and redshift to SNeIa ($H_0=74.03 \pm 1.42$ km s$^{-1}$ Mpc$^{-1}$) at $4.0$--$5.8\sigma$ significance~\cite{Verde:2019ivm}. 
Recently, other probes have been made in order to clarify the issues, but still without success: Gravitationally-lensed time delays \cite{Chen:2019,Wong:2019} from quasars analysed from the H0LiCOW collaboration \cite{Suyu:2016qxx} gives $H_0=73.3^{+1.7}_{-1.8}$ km s$^{-1}$ Mpc$^{-1}$, a $2.4\%$ with a $3.1\sigma$ tension with \textit{Planck 2018}. The Dark Energy Survey (DES) collaboration \cite{Abbott:2018jhe} has used SNeIa and Baryon Acoustic Oscillations (BAO) in an \textit{inverse distance ladder} method to obtain $H_0=67.8\pm1.3$ km s$^{-1}$ Mpc$^{-1}$ \cite{Macaulay:2018fxi}. And finally, upgrades in the analysis of water masers in $NGC4258$ \cite{Reid:2019tiq} have give us a value of $H_0=73.5\pm1.4$ km s$^{-1}$ Mpc$^{-1}$, a $4.2\sigma$ tension with \textit{Planck 2018}. A full compendium of \hubble constant estimations are detailed in Figure \ref{fig:H0-all}.

To understand the implications of this discrepancy it is, by far, the most severe problem the standard cosmology is facing now. Several solutions have been proposed to explain, solve or even alleviate such a tension. For example, in  \cite{Borhanian:2020vyr} has been studied independent measurement of the host galaxy’s redshift with the luminosity distance from GW to infer $H_0$. Also, the KBC void has been considered an interesting proposal to solve naturally the \hubble tension in Milgromian dynamics \cite{Haslbauer:2020xaa}. Although there are studies that consider this local feature not important on $H_0$ measurements \cite{Kenworthy:2019qwq}.

Even if some criticisms inherent to the distance ladder method, both for the procedure and for the existence of a local void \cite{Shanks:2018rka} have been raised. Nowadays the main scenario focuses on translating such observational tension in a tension between our description and understanding of both late and early time physics \cite{DiValentino:2020zio}.
From one part, the effect of the local structure on $H_0$ (so-called cosmic variance) has been thoroughly studied, as well as possible reassessments of its error budget. Furthermore, physics beyond the standard model has been also investigated, in the hope that this discrepancy  could reveal possible alternatives to the highly tuned cosmological constant and the yet-undetected dark matter. Confirm or dismiss this discrepancy is of extreme importance as it may point out to new (or missing) physics beyond the recombination era.

Currently, a panorama has been open to understand much deeper this tension issue: the multi-messenger Gravitational Wave (GW) astronomy. The first combined detection of GW and EM waves from the same source has brought light to analyze astrophysical \cite{Meszaros:2019xej} and cosmological \cite{Ezquiaga:2018btd,Corman:2020pyr} level at the same time. As for example, independent estimations can be obtained through the LIGO collaboration \cite{Abbott:2017xzu}, which value reported is $H_0=70.0^{+12.0}_{-8.0}$ km s$^{-1}$ Mpc$^{-1}$ from the detection of a binary neutron star inspiral \cite{2017PhRvL.119p1101A}. Dark siren detected in the first and second observing runs of LIGO and Virgo estimated  $H_0=68.0^{+14.0}_{-7.0}$ km s$^{-1}$ Mpc$^{-1}$ \cite{LIGOScientific:2018mvr}.
In comparison to other methods, this result from GW is settled in the middle. Unfortunately, as we can see from the error bars, this measurement cannot help to solve the tension, yet. Also, radio signals \cite{Hotokezaka:2018dfi} have been considered to compute a value of $H_0=68.9^{+4.7}_{-4.6}$ km s$^{-1}$ Mpc$^{-1}$, which is consistent with both \textit{Planck 2018} and local estimators.

Until the next era of surveys is being ready in order to set light on this issue, our main objective should be to find new probes (or new alternative ways to employ current astrophysical data) which might add information to the argument and be competitive with measurements for what concerns the feasible precision. In that regard, new physics via an alternative theory of gravity should describe gravitational phenomena in a very wide range of systems, from the cosmic scales to compact objects.  Hence, we can combine several observations to improve bounds on parameters.  In the era of multi-messenger astronomy, modelling GW  propagation is the new window in alternative theories of gravity and some of its inherent challenges, therefore in this paper we will study the possibility that late dark energy itself through a new EoS proposal can solve the \hubble tension in the late universe. We should mention that possibilities with an extra dark energy in the early universe has been considered \cite{Niedermann:2019olb}, moreover this involves first order transitions and many parameter assumptions to be fitted with the data. The aim of our study is slightly different: we will consider an exponential-like late dark energy parameterisation that in its first order
approximation at present cosmic time recovers the standard dark energy models, e.g. CPL \cite{Chevallier:2000qy,Linder:2007wa}, BA \cite{barboza2008parametric}, etc. and at higher order approximation can be compatible to be constrained using GW data to investigate how such \textit{corrections} affect the $H_0$ tension at late-time.

This paper is organised as follows: in Sec.\ref{sec:DE_test} we present our proposal for a new late dark energy EoS and how it can have cosmological dependencies on the luminosity distance. In Sec.\ref{sec:DS} we described the method to obtain
the \hubble constant using a dark sirens and the requirements needed from GW LIGO-Virgo catalogues. In Sec.\ref{sec:method} we introduce the methodology to deal with the GW LIGO-Virgo transient catalogues GWTC-1 and GWTC-2 and study our late dark energy proposal. In Sec.\ref{sec:results} we discuss the statistical results for the cosmology derived and finally, in Sec.\ref{sec:conclusions} we present our comments.

\section{Testing Hubble-Lema\^{i}tre with dark energy}
\label{sec:DE_test}

A possible way to describe the current observational evidences is to introduce dark energy, a
hypothetical fluid with negative pressure. Moreover, apart from this negativity property on the pressure, it is still unknown the \textit{nature} of this dark component. The simplest and standard explanation comes through the introduction of a positive cosmological constant, $\Lambda$, which does not evolve with the time, but carries out with itself two major problems: one of which is the vacuum catastrophe and the other is the cosmic coincidence problem. These issues motivate us to consider scenarios beyond the standard $\Lambda $-cosmology landscape. 

The simple scenario beyond the standard one, is the EoS $w(z)$-cosmology where
$w$ quantified the ratio of pressure $p$ to its density $\rho$, i.e. $w= p /\rho$. Of course, a simple recovery of $\Lambda$-cosmology can be obtained easily when $w=-1$. Any deviation from this value can give a different cosmology, moreover, we shall focus on
the alternative cosmologies in which the dark energy EoS is evolving with the Hubble flow.

Over the years, in the literature has been presented wide compendiums of $w(z)$ parameterisations as any arbitrary function of the redshift $z$ or the scale factor $a$ of the Friedmann-Lema\^{\i}tre-Robertson-Walker universe. The advantage of such idea gives a complete freedom to select any specific model of interest and test it with the surveys at hand to study whether that model is able to correctly describe the evolution
of the universe. In such a way, we can think this as an inverse evolving cosmology: we introduce a dark energy EoS to probe the expansion
history of the universe. This is one of the motivation of using late dark energy as a model to solve the \hubble tension. While possibilities with an extra dark energy in the early universe has been considered \cite{Niedermann:2019olb}, remember that in this era we are restricted by using the only data we have at hand now: CMB data, which assume a $\Lambda$CDM-cosmology. Moreover, a design of late dark energy that can be tested using independent measurements as GW sirens, can shed some light on the tension problem without theoretical dependencies.

The aim of the present study consist in propose an exponential late dark energy parameterisation that in its zero and 
first order approximation around present time ($z=0$) recovers $\Lambda$CDM and the CPL parameterisation, respectively. Furthermore, we allow its higher order approximation to study how such extended corrections can solve the \hubble tension. As it is standard, this kind of proposals emerge from a Taylor series inspiration. Let us start by proposing
 a late dark energy fluid with the following form: 
\begin{equation}
\label{eq:eos-general}
w(z)=w_{0}+w_{1}e^{\left( \frac{z}{1+z}\right)},
\end{equation}
where $w(z=0)=w_{0}+w_1$  is the current value of the dark energy EoS and $w_{1}$ is another free parameter. 
We can notice some interesting features of this proposal: 
\begin{enumerate}
\item The pivoting redshift can be found in order to uncorrelated the parameters $w_0$ and $w_1$ \cite{Yang:2018prh}.
\item By this approach we will get important information on how the non-linear terms in the parametrised EoS affect the viability of the cosmological model in higher-redshifts. 
\item And finally, in comparison to other dynamical late dark energy whose deals with divergencies at low and high $z$ on the Taylor series, our proposal by its construction includes a smoother evolution that can be controlled via the its first, second and third order corrections.
\end{enumerate}


\subsection{Cosmological dependencies}

At this point, it is interesting to have an insight on how our proposal (\ref{eq:eos-general}) might depend on the cosmological parameters, and in particular on $H_0$. First of all, one should note that most of the dependence comes from the luminosity distance definition
\begin{eqnarray}\label{eq:DL}
D_L(\Theta)=(1+z)\int_0^z{\frac{c\, dz'}{H_0 E(z',\Theta)}},
\end{eqnarray}
and $\Theta =\{w_0,w_1\}$ is the vector with the free cosmological parameters to be fitted and $E=H/H_0$.
To introduce our proposal (\ref{eq:eos-general}) 
we describe the expansion history by the following expression:
\begin{eqnarray}\label{eq:hubbleparam}
H^2(z) = H_0^2\ \left[ \Omega_m (1+z)^3 + \Omega_\Lambda \, f(z) \right],
\end{eqnarray}
where
\begin{equation}
f(z) = \exp \left( 3 \int_0^z \frac{1+w(z')}{1+z'} dz' \right).
\end{equation} 
This expression arise from the energy-momentum tensor conservation law for dark energy.
Notice that $\Lambda$CDM is a special case $w(z)=-1$, so $f(z)=1$. $\Omega_m$,  is the matter density parameter today; $\Omega_{\Lambda} = 1 - \Omega_m$, the dark energy density parameter; and $w(z)$ our proposal, which will be $-1$, in the case of dark energy as a cosmological constant. Our fiducial model is a flat cosmology, i.e. $\Omega_k =0$.

\section{$H_0$ with Dark Sirens}
\label{sec:DS}
 
\begin{figure}
\centering
\includegraphics[width=0.6\textwidth]{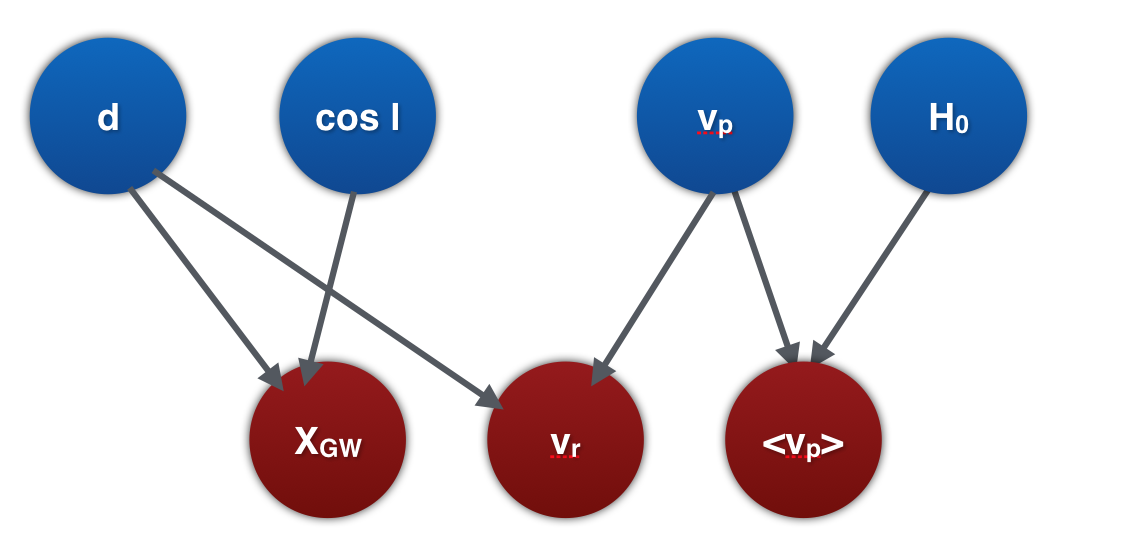}
\caption{Statistical relationships between the data and the parameters for a GW to compute $H_0$. Blue circles indicate parameters that require a prior. The red circles describe the measured data, which is conditioned on the analysis. Here we assume that we have measurements of the GW data, $x_{GW}$ (set of detectors), a recession velocity (that is, redshift) $v_r$, and the mean peculiar velocity $v_p$ in the galaxy's neighbourhood.
Arrows towards a node indicate that the conditional probability density for the node depends on the source parameters e.g. the conditional distribution of the observed GW data depends on the distance and inclination of the source (and additional parameters, here marginalized).}
\label{fig:GW_S}
\end{figure}

To compute the \hubble constant we need to estimate the velocity of the Hubble flow at the position of a galaxy at certain redshift, e.g. at $z = 0.01$. 
In order to have a a precise estimation (less than 2\%) it is require to correct peculiar velocities of the hosts of GW events, which correspond to a particular case for very low $z$ sources. This is due to the high signal-to-noise ratio which eventually generate a bias in the joint posteriors for set of events. Therefore, for a standard value of a peculiar velocity around $300$km/s for a GW host we consider $z=0.01$, since the contribution from peculiar velocity is comparable to the term related with the Hubble flow. Using an EM calibration of the \textit{cosmic distance ladder}, we can derive the distance to this galaxy using the Tully-Fisher relationship. Since in this relationship the method is still model-dependent, we estimate the velocity of the Hubble flow at the position of the galaxy considering corrections induced by local peculiar motions. Finally, we can constrain the value of the \hubble constant by using the distance and velocity distributions derived from the GW and EM data processed via Bayesian statistics.
Treating with GW observation is something else entirely. Its method behind allow us to completely jump over the distance ladder and obtain an independent measurement of the distance to cosmological objects. The simultaneous observation of gravitational and EM waves is crucial to this idea since we compare two things: the distance to an object, and the apparent velocity with which it is moving away from the observer. Measure the redshift $z$ of light is easy to do when we have an EM spectrum of a specific object. Moreover, with GW alone, we cannot do it — there is not enough structure in the spectrum to measure a $z$. 
Moreover, GW cannot give us standard candles \textit{per se}, since every one will have a different intrinsic gravitational luminosity, but 
we can work what that luminosity is if we take into account the way in which the source evolves.
The idea of using GW sources as \textit{standard sirens} was study by B. Schutz in 1986 \cite{Schutz:1986gp}. Since then, the idea been developed substantially since \cite{Holz:2005df}, given us a step forward to use Bayesian analysis. For example, in \cite{Gray:2019ksv} it is described an analysis of mock data using binary neutron star mergers to recover an unbiased estimate of $H_0$.
In Figure \ref{fig:GW_S} we illustrate the statistical relationships between the data and the parameters to compute $H_0$ using sirens.

Moreover, for specific phenomena as binary black hole (BBH) mergers, we do not expected to have EM counterparts but
they can determine the luminosity distance to their hosts independently of the cosmic distance ladder method. These are usually referred to as \textit{dark sirens}. 
The GW standard dark siren measurement is broadly consistent with other measurements \cite{LIGOScientific:2019zcs}. If we combine the information from multiple detections, we can improve the accuracy reaching about 1\% with $O$(100) detections in the next decade. Moreover, the BBH sources with no expected EM counterparts can be used with the method outlined in  \cite{Schutz:1986gp}: with a database of potential host galaxies identified in a galaxy catalog for each detection, we can build up data by a process of statistical cross-correlation. These results established the path to cosmology using GW observations with and without transient EM counterparts. The first results of the implementation of this method was presented in \cite{DelPozzo:2011vcw} using a set of simulations with 5\% estimate on $H_0$ in a idealised situation where nearby events and complete galaxy catalogs have similar results for a third-generation detector \cite{LIGOScientific:2021inr}.

According to this, GW from compact binary coalescences, being standard sirens, can be used to calibrate distances to SNeIa, when both occur in the same galaxy or galaxy cluster \cite{Gupta:2019okl}.
Furthermore, it can be possible to identify their potential host galaxies either with the help of a galaxy catalog or looking coalescence effects from stellar-mass BBH \cite{Nishizawa:2016ood}. As it is standard in the statistical analysis, we require taking into account the sources of systematic errors in this method, which arises from the incompleteness
of the available galaxy catalogs. Nonetheless, LIGO and Virgo at their design sensitivity could constrain $H_0$
with dark sirens to 5\% accuracy with $\sim250$ detections \cite{Nair:2018ign}. It is in this direction that an independent estimation from the multi-messenger detection of a binary neutron star inspiral \cite{2017PhRvL.119p1101A} has obtained $H_0=70.0^{+12.0}_{-8.0}$ km s$^{-1}$ Mpc$^{-1}$. Additionally, in our study we include GW searches for coalescing compact binaries from GW transient catalogs GWTC-1 \cite{LIGOScientific:2018mvr} and  GWTC-2 \cite{Abbott:2020niy}, which allow us to include larger solar masses and coalescence effects.

\section{Methodology}
\label{sec:method}

LIGO and Virgo have published two compact binary mergers found during the third observing run: GW190412~\cite{LIGOScientific:2020stg} and
~\cite{Abbott:2020khf}, which are exceptional due to their large
mass-asymmetry, with mass ratios around $\sim 3$ and $\sim 9$, respectively. This kind of systems have led to the detection of subdominant spherical harmonic modes beyond the quadrupole mode. Therefore, higher modes will be important in the parameter estimation of asymmetric
binaries, e.g. of the luminosity distance $D_L$ and orbital inclination $\iota$. All these measurements can be attached with the new runs that includes upgrades to GW LIGO-Virgo transient catalogs GWTC-1 and GWTC-2, with sensitive of only 10\% of contamination fraction.

In this work we are interested in compute local values of $H_0$ with very distant sources. Consider that, if sources are near (e.g. at $D_L\lesssim 100\,\rm Mpc)$, its measurements could be imprecise due to the bias from peculiar velocities $v_p$ of host galaxies. In this regime, any proposal mimicking $\Lambda$CDM can work. Therefore, we will consider our dark energy proposal distributed uniformly in a volume up to a redshift of $z=2.3$, including along GW measurements, supernovae and observational Hubble data.

\subsection{Distance-ladder measurements}

\begin{itemize}
\item Pantheon SNeIa compilation \cite{Scolnic:2017caz}: contains 1048 SNeIa distribute in a redshift range of $0.01<z<2.26$. The constraining power of this supernovae is due that can be used as standarizable candles. Its implementation is via 
\begin{eqnarray}
\mathcal{F}(z,\Theta)_\text{theo}=5\log_{10}\left[D_L(z,\Theta)\right]+\mu_0,
\end{eqnarray}
where $D_L$ is the luminosity distance given by (\ref{eq:DL}) and $\mu_0$ is the nuisance parameters of the distance estimator. We can write $\Delta\mathcal{F}(\Theta)=\mathcal{F}_{\text{theo}}-\mathcal{F}_{\text{obs}}$, using for this purpose the  distance modulus $\mathcal{F}_{\text{obs}}$ associated with the observed magnitude and marginalise $\chi_{SN}^2$ with respect to $\mu_0$
\begin{eqnarray}
\chi_{SN}^2(\Theta)&=&\left(\Delta\mathcal{F}(\Theta)\right)^{T}\cdotp C_{SN}^{-1}\cdotp \Delta\mathcal{F}(\Theta)+\ln{\frac{S}{2\pi}} \nonumber\\
&&-\frac{k^2(\Theta)}{S},
\end{eqnarray}
where $C_{SN}$ is the total covariance matrix and $S$ is the sum of all entries of $C_{SN}^{-1}$, weighed by a covariance matrix as 
$k(\Theta)={\left(\Delta\mathcal{F}(\Theta)\right)^{T}\cdotp C_{SN}^{-1}}.$
\\
\item Observational Hubble data (CC) \cite{Moresco:2016mzx}: contain $l=51$ measurements in the redshift range $0.07< z < 2.0$. In this sample, 31 data points correspond to passive galaxies and 20 data points are estimated from BAO data under a $\Lambda$CDM prior. 
Our theoretical setting to construct $\chi_H^2$ is given by 
\begin{eqnarray}
\chi_H^2=\sum_{i=1}^{l}\frac{\left[H\left(z_i,\mathbf{x}\right)-H_{\textit{obs}}(z_i)\right]^2}{\sigma^2_H(z_i)},
\end{eqnarray}
where $H_{\textit{obs}}(z_i)$ is the observed value at $z_i$, $\sigma_H(z_i)$ are the observational errors, and $H\left(z_i,\mathbf{x}\right)$ is the value of a theoretical $H$ for the same $z_i$ with the parameter vector $\mathbf{x}$. 

\end{itemize}

\subsection{Gravitational waves measurements}

\begin{itemize}

\item Gravitational-Wave Transient Catalog GWTC-1 \cite{abbott2019gwtc}.
Includes $N_{\text{GWTC-1}}=11$ confident detection events in redshift range of $z \in [0.01,0.49]$ from GW searches for coalescing compact binaries, all of them with masses $M > 1 M_{\odot}$. These events observed by LIGO \footnote{\url{https://www.ligo.caltech.edu}} and Virgo\footnote{\url{http://www.virgo.infn.it}} come from the first and second observing runs. Eleven events are reported along with their observational siren distances $D_S$ in Mpc and redshifts $z$ and other several observational variables, all of them with their corresponding $90 \%$ credible variances. 
Its statistics can be described by
\begin{widetext}
\begin{eqnarray}
	\chi_{\text{GWTC-1}}^2 = 
	 \sum_{j=1}^{N_\text{GWTC-1}} \frac{(D_{S}(z_i, h, \Omega_M,w_0,w_1, n, \Xi_0)-D_{S \, \text{obs, GWTC-1}}(z_i))^2}{\sigma_{j \, \text{obs, GWTC-1}}^2},
\end{eqnarray}
\end{widetext}
where the quantities $n$ and $\Xi_0$ are both positive and defined as \cite{Mitra:2020vzq}
\begin{eqnarray}
\Xi_0 &=& \lim_{z \to \infty} \frac{M^*(0)}{M^*(z)}, \\
n &\approx& \frac{\alpha_{M_0}}{2(\Xi_0-1)},
\end{eqnarray}
with $M*=m^{2}_{\text{pl}}/\sqrt{G}$ as the effective time-dependent Planck mass, where $G$ is the gravitational strength. Also, this catalog
includes variances in $z$, therefore the squared variances are given by
\begin{equation}
	\sigma_{j \, \text{obs, GWTC-1}}^2 = \sigma_{j \, D_S \text{obs, GWTC-1}}^2 + \sigma_{j \, z \text{obs, GWTC-1}}^2,
\end{equation}
where the siren luminosity distance is proposed as
\begin{equation}
D_{S(z)} = D_L(z) \exp \left( - \int_{0}^{z} \frac{\delta(z')}{1+z'} dz' \right),
\end{equation}
with
\begin{equation}
\delta_1 (z) = \frac{n (1- \Xi_0)}{1-\Xi_0+\Xi_0(1+z)^n}.
\end{equation}
The squared variances of the siren distance per redshift are already added. 

\item Gravitational-Wave Transient Catalog GWTC-2 \cite{Abbott:2020niy}. This catalogue includes $N_{\text{GWTC-2}}=39$ GW events, with less than $10\%$ of contamination fraction in a range of $z \in [0.03,0.8]$ from the first half of the third observable run. 
As in the latter catalogue, the statistics can be described by
\begin{widetext}
\begin{equation}
	\chi_{\text{GWTC-2}}^2 = \sum_{j=1}^{N_\text{GWTC-2}} \frac{(D_{S}(z_i, h, \Omega_m,w_0,w_1, n, \Xi_0)-D_{S \, \text{obs, GWTC-2}}(z_i))^2}{\sigma_{j \, \text{obs, GWTC-2}}^2},
\end{equation}
\end{widetext}
with variances in $z$ as
\begin{equation}
	\sigma_{j \, \text{obs, GWTC-2}}^2 = \sigma_{j \, D_S \text{obs, GWTC-2}}^2 + \sigma_{j \, z \text{obs, GWTC-2}}^2.
\end{equation}
The squared variances of both siren distances and redshift are added up. Due that these events have less than $10\%$ of contamination fraction, the best fit values and their confidence regions might be biased, even though some bias can be controlled when using the distance ladder measurements described above.
\end{itemize}


\section{Results}
\label{sec:results}

\begin{figure}
	\centering
	\includegraphics[scale=0.5]{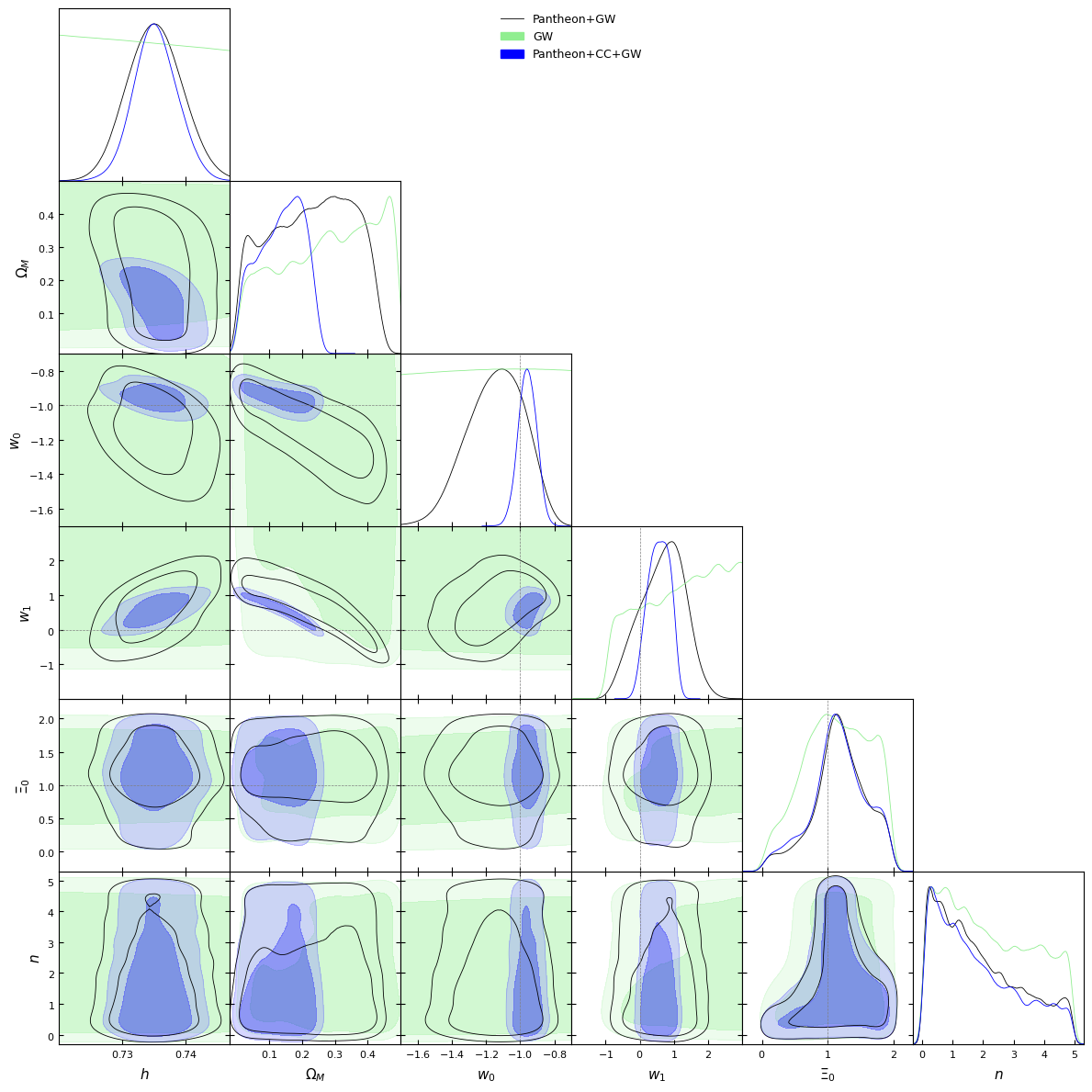}
	\caption{For our model (\ref{eq:eos-general}), we first constrain the cosmological parameters ($h$, $\Omega_m$, $w_0$, $w_1$, $\Xi_0$ and $n$) using the datasets Pantheon+GW (including both GWTC-1 and GWTC-2) (black solid lines), GW alone (green C.L.) and Pantheon+CC+GW (blue C.L.), respectively. The best-fit values of the parameters for each dataset are reported in Table \ref{tab:results}.}
	\label{fig:results}
\end{figure}

We constrain our parameterisation (\ref{eq:eos-general}) with distance ladder measurements, such as Pantheon and CC described above. Also, we consider another dark energy models (see Table \ref{tab:models}) to compare with our proposal (\ref{eq:eos-general}).
Afterwards, we employed the best fit values obtained (see Table \ref{tab:results}) to generated constraints on dark sirens in the transient catalog GWTC-1 and GWTC-2.  In Figure \ref{fig:results} we present the comparisons between the constraining results of the data samples after the inclusion of the GW data to distance ladder measurements, where in particular, we show the
1-dimensional marginalised posterior distributions for the parameters of our model $(w_0, w_1)$ and the ones derived from the GW (${\Xi_0, n}$), as well as the 2-dimensional contour plots (C.L.) between the combinations Pantheon+GW, GW and Pantheon+CC+GW.
In the following we describe the effects of GW on the (\ref{eq:eos-general}) corresponding to different observational data samples. 

In Table \ref{eq:eos-general}, we present five of the most popular dark energy parameterisations in the literature (a study of these models with current data was presented in \cite{Escamilla-Rivera:2016qwv}). Also, we include in the analysis of our model (\ref{eq:eos-general}) and the $\Lambda$CDM model. In the second column of this table, we described the data combinations employed to perform the fit of the cosmological parameters.
One can clearly notice that the treatment of GW solely for $\Lambda$CDM and the five dark energy models shows a \hubble constant value in the periphery with the $H_0$ inferred from the fluctuation spectrum of the CMB, while in comparison to the other proposals, our model gives a value of $H_0=68.0\pm 0.077$ km s$^{-1}$ Mpc$^{-1}$, closer to the dark siren  detected in the first and second observing runs of LIGO and Virgo estimated  $H_0=68.0^{+14.0}_{-7.0}$ km s$^{-1}$ Mpc$^{-1}$. The error bars seems to decrease when we consider, along with the GWTC-1 catalogue, the GWTC-2, this consequence is directly connect to the fact that we are dealing with a sensitivity of 10\% of contamination fraction in the catalogues and their resolution allow us to measure $H_0$ with dark sirens to within 2$\sigma$. We also note that the matter density parameter $\Omega_{m0}$, for GW alone is better constrained compared to the Planck's estimation and the inclusion of GW to SN and CC improves the parameter space.

\begin{table*}
\caption{Description of the dynamical parameterisations used in this work. In the first column we denote the model and the reference and in the second column their expression for $w(z)$. This table complements the analysis in Table \ref{tab:results}.
 	}
	{\centering
	\renewcommand{\arraystretch}{1.}
	\begin{tabular}{|c|c|c|c|c|c|c|c|}
		\hline
		Model & Parameterisation  \\ \hline
		Our Model (\ref{eq:eos-general}) & $w(z)=w_{0}+w_{1}e^{\left( \frac{z}{1+z}\right)}.$
		 \\ \hline
		Lambda Cold Dark Matter ($\Lambda$CDM)  & $w=-1.$
		 \\ \hline
		 Chevallier-Polarski-Linder (CPL)  \cite{Chevallier:2000qy,Linder:2007wa}  & $w(z) = w_0 + \left(\frac{z}{1+z}\right) w_1.$
		 \\ \hline
		 	Barboza-Alcaniz (BA)  \cite{barboza2008parametric} & $w(z)=w_0 + \frac{z(1+z)}{1+z^2} w_1.$
		 \\ \hline
		 	Linear-redshift (LR) \cite{PhysRevD.64.123527}  & $w(z)=w_0-w_1 z.$
		 \\ \hline
		 	Low Correlation (LC) \cite{wang2008figure}  & $w(z) = \frac{(-z+z_c)w_0+z(1+z_c)w_c}{(1+z)z_c}.$
		 \\ \hline
		 	Wetterich (WP) \cite{wetterich2004phenomenological} & $w(z) = \frac{w_0}{(1+w_1 \ln(1+z))^2}.$
		 \\ \hline
		 	\end{tabular}
	\label{tab:models}}
\end{table*}

\begin{table*}
\caption{Best fits cosmological parameters obtained using GW, Pantheon (SN) and CC. We include the $\Lambda$CDM model, our proposed model (\ref{eq:eos-general}) and five dark energy bidimensional parameterisations: CPL \cite{Chevallier:2000qy,Linder:2007wa}, BA \cite{barboza2008parametric}, LR \cite{PhysRevD.64.123527}, LC \cite{wang2008figure} and WP  \cite{wetterich2004phenomenological}.
	 A full review of these models with late-universe data is presented in \cite{Escamilla-Rivera:2016qwv}, and references therein. All these values are at 2$\sigma$ of uncertainties C.L.
 	}
	{\centering
	\resizebox{\textwidth}{!}{  
	\renewcommand{\arraystretch}{0.7}
	\begin{tabular}{|c|c|c|c|c|c|c|c|}
		\hline
		Model & Data & $h$ & $\Omega_m$ & $w_0$ & $w_1$ & $\Xi_0$ & $n$  \\ \hline
		\multirow{3}{*}{(\ref{eq:eos-general})} & GW & $0.680\pm 0.077$ & $0.28^{+0.21}_{-0.14}$ & $-1.4^{+2.2}_{-1.6}$ & $-0.1^{+3.8}_{-4.5}$ & $1.06^{+0.49}_{-0.56}$ & $2.17^{+0.87}_{-2.1}$ \\
		& SN+GW & $0.7344\pm 0.004$ & $0.299^{+0.12}_{-0.04}$ & $-1.21^{+0.17}_{-0.15}$ & $0.6^{+2.3}_{-1.0}$ & $1.18^{+0.24}_{-0.28}$ & $1.8^{+1.1}_{-1.8}$ \\
		& SN+CC+GW & $0.7348\pm 0.004$ & $0.216^{+0.024}_{-0.018}$ & $-1.043^{+0.074}_{-0.083}$ & $0.89\pm 0.74$ & $1.17^{+0.24}_{-0.29}$ & $1.77^{+0.69}_{-1.8}$ \\ \hline
		\multirow{3}{*}{$\Lambda$CDM} & GW & $0.690\pm 0.073$ & $0.26\pm 0.14$ & - & - & $1.00\pm 0.40$ & $1.93^{+0.77}_{-1.9}$ \\
		& SN+GW & $0.728\pm 0.002$ & $0.286\pm 0.013$ & - & - & $1.19^{+0.24}_{-0.30}$ & $1.8^{+1.2}_{-1.8}$ \\
		& SN+CC+GW & $0.733\pm 0.002$ & $0.245\pm 0.007$ & - & - & $1.17^{+0.25}_{-0.28}$ & $1.7^{+1.2}_{-1.7}$ \\ \hline
		\multirow{3}{*}{CPL  } & GW & $0.698\pm 0.078$ & $0.286^{+0.21}_{-0.079}$ & $-1.4^{+2.2}_{-1.7}$ & $-0.1^{+2.9}_{-4.6}$ & $1.04^{+0.48}_{-0.57}$ & $2.2^{+1.5}_{-2.1}$ \\
		& SN+GW & $0.734\pm 0.004$ & $0.301^{+0.13}_{-0.046}$ & $-1.18\pm 0.17$ & $0.08^{+1.7}_{-0.62}$ & $1.18^{+0.24}_{-0.30}$ & $1.8^{+1.1}_{-1.8}$ \\
		& SN+CC+GW & $0.734\pm 0.003$ & $0.169^{+0.077}_{-0.030}$ & $-0.951^{+0.054}_{-0.074}$ & $0.61^{+0.46}_{-0.18}$ & $1.16\pm 0.34$ & $1.75^{+0.98}_{-1.8}$ \\ \hline
		\multirow{3}{*}{BA } & GW & $0.698\pm 0.078$ & $0.281^{+0.21}_{-0.093}$ & $-1.4^{+2.3}_{-1.6}$ & $-0.2^{+2.1}_{-4.4}$ & $1.06^{+0.51}_{-0.58}$ & $2.2^{+1.1}_{-2.2}$ \\
		& SN+GW & $0.7329\pm 0.004$ & $0.332^{+0.11}_{-0.033}$ & $-1.20^{+0.15}_{-0.19}$ & $-0.34^{+1.3}_{-0.42}$ & $1.18^{+0.25}_{-0.29}$ & $1.8^{+1.6}_{-1.8}$ \\
		& SN+CC+GW & $0.7331\pm 0.003$ & $0.157^{+0.088}_{-0.037}$ & $-0.913^{+0.056}_{-0.092}$ & $0.324^{+0.21}_{-0.056}$ & $1.17^{+0.24}_{-0.29}$ & $1.8^{+1.0}_{-1.8}$ \\ \hline
		\multirow{3}{*}{LR } & GW & $0.698^{+0.085}_{-0.075}$ & $0.282^{+0.21}_{-0.092}$ & $-1.4^{+2.3}_{-1.6}$ & $0.2\pm 2.9$ & $1.07\pm 0.49$ & $2.2^{+1.2}_{-2.1}$ \\
		& SN+GW & $0.7328\pm 0.004$ & $0.324^{+0.12}_{-0.034}$ & $-1.18^{+0.15}_{-0.20}$ & $0.33^{+0.48}_{-1.4}$ & $1.17^{+0.25}_{-0.30}$ & $1.75^{+0.69}_{-1.8}$ \\
		& SN+CC+GW & $0.7314\pm 0.003$ & $0.211^{+0.043}_{-0.020}$ & $-0.937^{+0.044}_{-0.056}$ & $-0.098^{+0.095}_{-0.24}$ & $1.16^{+0.25}_{-0.28}$ & $1.77^{+0.94}_{-1.8}$ \\ \hline
		\multirow{3}{*}{LC } & GW & $0.691^{+0.098}_{-0.077}$ & $0.28^{+0.21}_{-0.13}$ & $-0.9^{+1.7}_{-3.3}$ & $-0.8^{+1.8}_{-3.4}$ & $1.12\pm 0.48$ & $2.18^{+0.88}_{-2.1}$ \\
		& SN+GW & $0.7333\pm 0.004$ & $0.309^{+0.13}_{-0.044}$ & $-1.18\pm 0.17$ & $-1.21^{+0.77}_{-0.32}$ & $1.17^{+0.25}_{-0.29}$ & $1.8^{+1.4}_{-1.8}$ \\
		& SN+CC+GW & $0.7337\pm 0.003$ & $0.172^{+0.075}_{-0.029}$ & $-0.953^{+0.051}_{-0.069}$ & $-0.75^{+0.18}_{-0.15}$ & $1.17^{+0.26}_{-0.29}$ & $1.74^{+0.61}_{-1.8}$ \\ \hline
		\multirow{3}{*}{WP} & GW & $0.695\pm 0.079$ & $0.27^{+0.22}_{-0.12}$ & $-1.1\pm 2.1$ & $2.3^{+2.6}_{-1.7}$ & $1.14^{+0.68}_{-0.44}$ & $2.2^{+1.6}_{-2.2}$ \\
		& SN+GW & $0.7351\pm 0.004$ & $0.22\pm 0.12$ & $-1.14^{+0.19}_{-0.15}$ & $0.67^{+0.76}_{-0.58}$ & $1.17\pm 0.33$ & $1.8^{+1.0}_{-1.8}$ \\
		& SN+CC+GW & $0.7353\pm 0.003$ & $0.134^{+0.086}_{-0.056}$ & $-0.958\pm 0.054$ & $0.58^{+0.36}_{-0.28}$ & $1.17^{+0.25}_{-0.29}$ & $1.8^{+1.1}_{-1.8}$ \\ \hline
	\end{tabular}
	}
		\label{tab:results}}
\end{table*}

\section{Conclusions}
\label{sec:conclusions}

We have demonstrated a possibility of measuring the \hubble constant to $\sim$2$\sigma$-level precision using dark sirens with the upgrades of the LIGO and Virgo detectors through the transient catalogs GWTC-1 and GWTC-2.  
Although the standard model $\Lambda$CDM has still some advantage in comparison other dark energy parameterisations (see Table \ref{tab:results}), GWs have opened the door to update the analysis previously done with supernovae and Hubble data. In fact, from the \hubble constant fit derived, our proposal (\ref{eq:eos-general}) offers the opportunity to measure $H_0$ with dark sirens GW to within 2$\sigma$, the accuracy required to resolve the \hubble tension, to obtain a value of $H_0 = 68.0 \pm 0.077$ km s$^{-1}$ Mpc$^{-1}$, closer to the dark siren  detected in the first and second observing runs of LIGO and Virgo estimated.

Future GW data might give us more clues to discover the actual model of dark energy, whether it continues to be $\Lambda$CDM or another one. The analyses done in this work will be improved with future numerous and accurate data and establish the preference of GR over the siren distance - luminosity distance non equality coming from modified gravity models \cite{belgacem2019testing}. This analysis will be reported elsewhere.  

In our opinion, only a complete and coherent theory of dark energy selected by high level of precision and control
of systematic in observations can infer the correct cosmological paradigm. We expect that the ideas presented in this work could give further 
insight into the \hubble tension issue; but that we shall know from the future work.


\begin{acknowledgments}
 CE-R acknowledges the \textit{Royal Astronomical Society} as FRAS 10147, supported by DGAPA-PAPIIT-UNAM Projects IA100220 and TA100122.
The simulations were performed in Centro de c\'omputo Tochtli-ICN-UNAM as part of \href{https://www.nucleares.unam.mx/CosmoNag}{CosmoNag}.
\end{acknowledgments}


\bibliographystyle{unsrt}
\bibliography{biblio} 

\end{document}